\documentclass[reprint,
amsmath,
hyperref,
aps, prstab
]{revtex4-1}

\usepackage{graphicx}
\usepackage{dcolumn}
\usepackage{bm}

\begin{document}


\title{Dependence of the residual surface resistance of superconducting RF cavities on the cooling dynamics around $T_\mathrm{c}$}

\author{A. Romanenko}
\email{aroman@fnal.gov}
\author{A. Grassellino}
\email{annag@fnal.gov}
\author{O. Melnychuk}
\author{D. A. Sergatskov}
\affiliation{Fermi National Accelerator Laboratory, Batavia, IL 60510, USA}%

\date{\today}
             
\begin{abstract}
We report a strong effect of the cooling dynamics through $T_\mathrm{c}$ on the amount of trapped external magnetic flux in superconducting niobium cavities. The effect is similar for fine grain and single crystal niobium and all surface treatments including electropolishing with and without 120$^\circ$C baking and nitrogen doping. Direct magnetic field measurements on the cavity walls show that the effect stems from changes in the flux trapping efficiency: slow cooling leads to almost complete flux trapping and higher residual resistance while fast cooling leads to the much more efficient flux expulsion and lower residual resistance. 
\end{abstract}

\maketitle

\section{Introduction}
Trapped magnetic flux represents one of the known contributors to the residual resistance $R_\mathrm{res}$ of superconducting radio frequency (SRF) niobium cavities~\cite{Hasan_book2}. Experiments showed that $R_\mathrm{res}$ due to trapped flux also increases with the magnitude of the RF field on the cavity surface~\cite{Benvenuti_PhysicaC_1999}, and can therefore have a significantly negative impact on the intrinsic cavity quality factor $Q_0$ at medium accelerating fields.

For this reason, minimization of trapped flux in niobium cavities has recently been a topic of particular interest, especially in light of its potential impact on cryogenic costs of high duty factor accelerators, i.e. LCLS-II, ERLs, and an XFEL upgrade. Studies at HZB showed that the details of the cooling procedure affect the amount of trapped flux and thus its associated additional residual surface resistance~\cite{Kugeler_IPAC_2013, Vogt_PRST_2013}. Based on the possible interpretation of the results, two main mechanisms of improvement of cavity performance (due to reduction of the amount of trapped magnetic flux) have been suggested: 1) slow cooling through transition temperature~\cite{Kugeler_IPAC_2013, Vogt_PRST_2013}; 2) reduction of thermocurrents, which are enabled by bimetal titanium-niobium junctions in dressed cavities, by minimizing temperature gradients~\cite{Vogt_PRST_2013}. Following HZB results, Cornell has also recommended a slow cool-down procedure based on the interpretation of their horizontal test results~\cite{Valles_IPAC_2013}. However, the physical mechanism of the effect has not been clearly established.

To illuminate the mechanism, it is very important to understand if the effect is specific to dressed cavities, or if it is a generic effect present also in bare niobium cavities. Furthermore, the effect of different surface treatments has to be understood as well. To do so, it is crucial to perform direct magnetic field measurements on the cavity in the cryostat to correspond with the RF measurements, which is the key component of our work.

\begin{figure*}[htb]
\includegraphics[width=\linewidth]{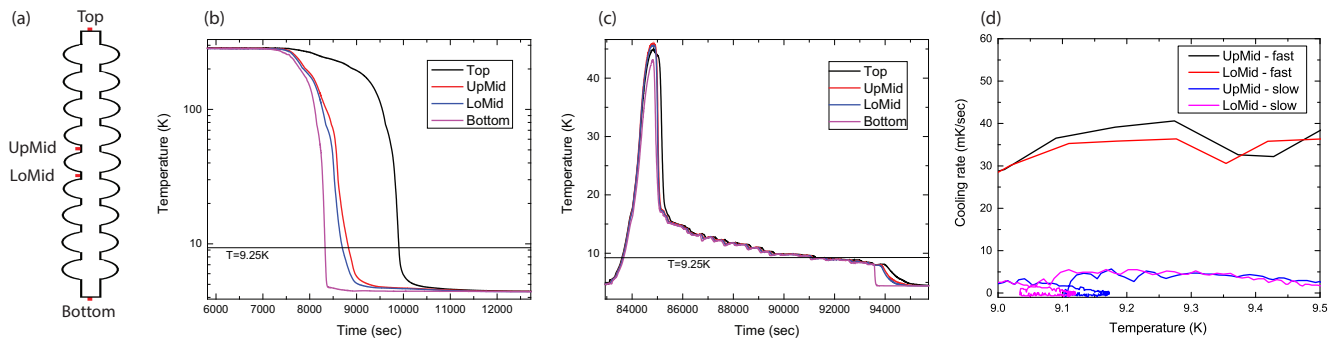}
\caption{\label{fig:Cooling}Example of: a) temperature sensors positions and readings for: b) fast and c) slow cooling procedures corresponding to one of the nine cell tests; d) corresponding cooling rates around $T_\mathrm{c}$ for UpMid and LoMid sensors.}
\end{figure*}

In this paper we present a set of systematic vertical test stand measurements on \emph{bare} single and nine cell 1.3 GHz TESLA elliptical shape SRF cavities made of fine grain ($\sim50\mu$m) and single crystal RRR$\sim$300 niobium, which reveal a strong effect of cooling protocol through $T_\mathrm{c}$ on the residual resistance: a slow cooling leads to higher residual resistance than the fast cooling. Meissner effect measurements directly on cavity walls show that the increase in the residual resistance is due to the smaller flux expulsion (larger flux trapping) at slow cooling rates, which provides an alternative explanation for the findings in Ref.~\onlinecite{Vogt_PRST_2013}.

\section{Experimental procedure}
We have performed all our measurements at the vertical test stand at Fermilab. Temperature during the experiments was continuously monitored at several locations in the cryostat by four Cernox thermometers attached to the outside cavity walls. RF measurements were performed using the standard phase-lock technique in the temperature range between 2~K and 1.5~K. Surface resistance $R_\mathrm{s}$ measured at the lowest temperature $T\sim$1.5~K was very close to the residual resistance $R_\mathrm{res}$ as was reconfirmed by the explicit deconvolution following our original procedure~\cite{Romanenko_Rs_B_APL_2013}. Thus in what follows we use terms $R_\mathrm{res}$ and $R_\mathrm{s}(T\leq1.5$~K) interchangeably. 

In order to prevent the possible occurrence of thermal currents we used additional measures to improve the electrical insulation of cavities from the supporting fixtures, e.g. kapton tape between the cavity and stainless steel holding fixtures or G10 holding fixtures instead of metal. Thus our work is not designed to address the effect of thermal currents but rather any intrinsic effects of the cooling dynamics itself.

Typical fast and slow cooling procedures we used are shown in Fig.~\ref{fig:Cooling}. In the case of the fast cooling various temperature differences up to 200~K can be present across the cavity (maximum depends on the starting temperature), and cooling rates through $T_\mathrm{c}$ are of the order of 30-40~mK/sec.  Slow cooling is very uniform with temperature differences on the order of 0.1~K and cooling rate through $T_\mathrm{c}$ of 2-5~mK/sec. 

\begin{figure*}[htb]
\includegraphics[width=0.7\linewidth]{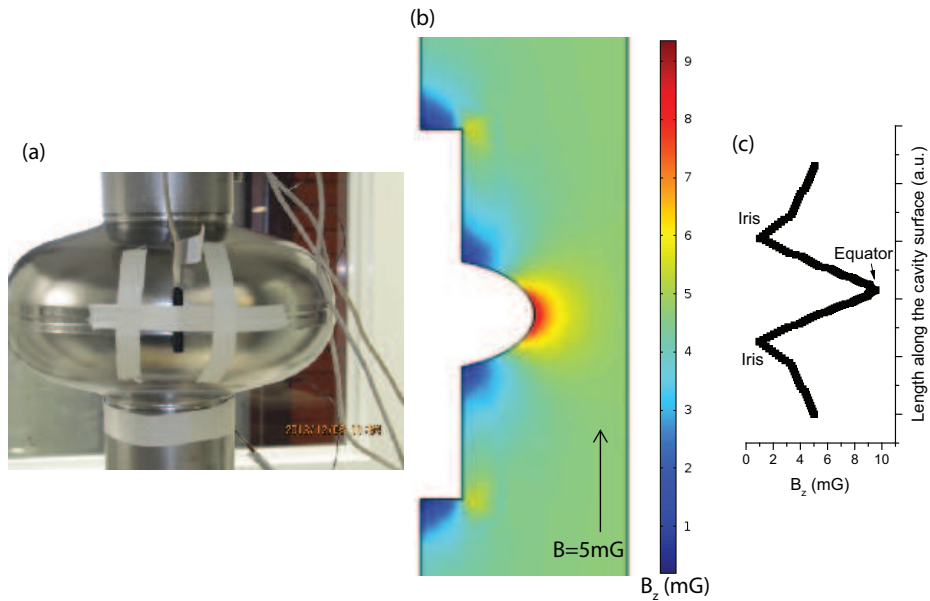}
\caption{\label{fig:Meissner}(a) Fluxgate magnetometer placement at the 1-cell cavity equator; (b) simulated distribution of the magnetic field after field cooling in the remnant field of 5~mG in vertical direction; (c) magnetic field amplitude over the cavity surface.}
\end{figure*}

For magnetic field measurements we used Bartington cryogenic Mag-01H single-axis fluxgate magnetometers attached to the outside cavity walls. Similar approach was originally implemented in Ref.~\onlinecite{Benvenuti_PhysicaC_1999} but using Hall probes. Fluxgate magnetometers were mounted in the vertical orientation to measure the magnetic field component parallel to the vertical symmetry axis of the cryostat unless specified otherwise in the text. An example of the magnetic probe placement on a 1-cell cavity is shown in Fig.~\ref{fig:Meissner}(a). Depending on the flux trapping efficiency, the transition to the Meissner state should lead to the expulsion of the magnetic flux from cavity walls and thus to an increase in the magnetic field amplitude measured right outside. If some of the flux remains trapped, the expulsion is smaller and the field outside changes less. COMSOL simulations assuming remnant field of $B=5$~mG in the axial direction are shown in Fig.~\ref{fig:Meissner}(b)-(c) and demonstrate that for the case of no trapping the magnetic field at the equator should be enhanced by about a factor of 2 when cavity is fully superconducting. Magnetic field measurements were taken upon both cool-down and warm-up through $T_\mathrm{c}$, providing full information regarding the magnetic field environment and the efficiency of flux expulsion.

\section{Results}

We have investigated cavities prepared by different state-of-the-art processing methods as described below.

\subsection{Electropolishing without 120$^\circ$C baking}

\begin{figure*}[htb]
\includegraphics[width=\linewidth]{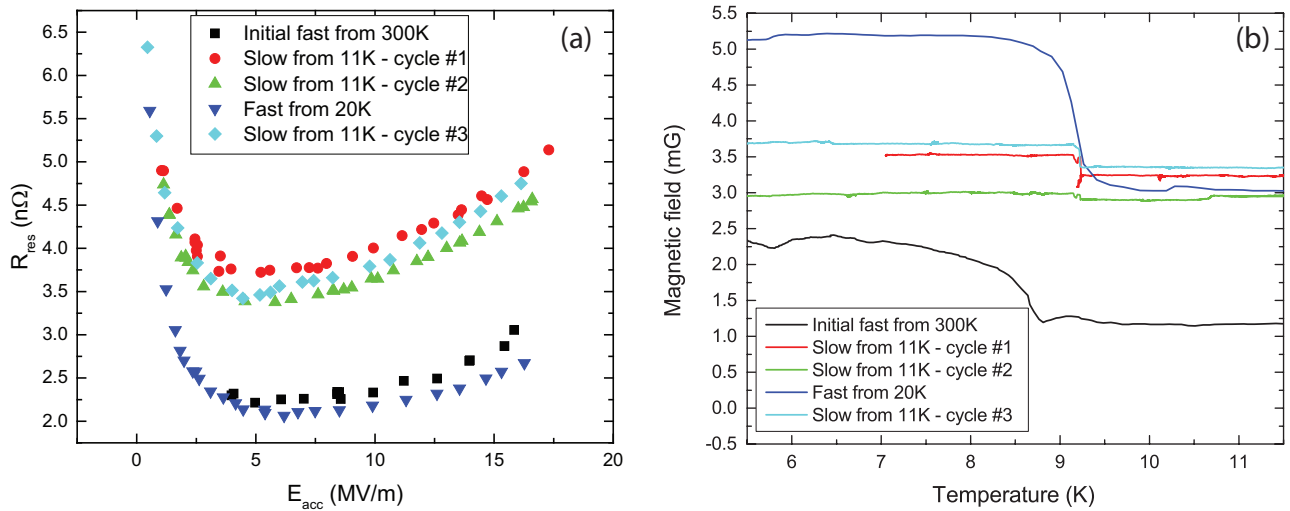}
\caption{\label{fig:SGEP}(a) Residual resistance of the single crystal 1-cell electropolished cavity after different cooling cycles; (b) corresponding magnetic field data at the equator during cool-down.}
\end{figure*}

\begin{figure}[htb]
\includegraphics[width=\linewidth]{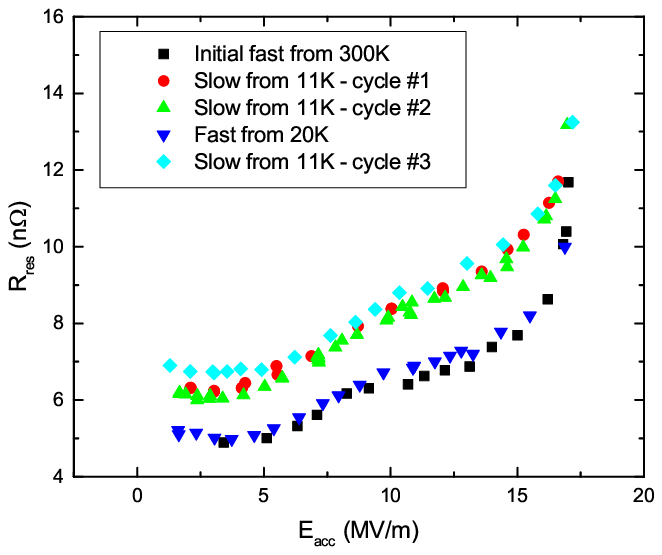}
\caption{\label{fig:FGEP}(a) Residual resistance of the fine grain 1-cell electropolished cavity after fast and slow cooling cycles.}
\end{figure}

In this experiment we used fine and single crystal 1-cell cavities after bulk electropolishing with about 120~$\mu$m of material removed. Cavities were mounted on the same test stand and magnetic probes were mounted on the equator of each cavity. Initial fast cooling from 300~K was followed by two cycles of warm-up to 11~K and slow cooling back to 4.2~K, warm-up to 20~K and a fast cooling to 4.2~K, and finally a warm-up and slow cooling from 11~K down to 4.2~K. Each of the cool-downs was followed by a full RF test.

Residual resistance $R_\mathrm{res}$ of the single crystal cavity as a function of $E_\mathrm{acc}$ measured at $T<1.5$~K is shown in Fig.~\ref{fig:SGEP}(a). The value of $R_\mathrm{res}$ was clearly and reproducibly dependent upon if the fast or slow cool-down had been performed. In particular, fast cool-downs from 300~K and 20~K (black squares and blue triangles) lead to similar $R_\mathrm{res}$ about 2~n$\Omega$ \emph{lower} than that after all slow cool-downs from 11~K (red circles, green triangles, and magenta diamonds). The value of the strongly temperature dependent BCS surface resistance remained approximately the same for all the RF tests. 

The magnetic probe readings around $T_\mathrm{c}$ are shown in Fig.~\ref{fig:SGEP}(b). The magnitude of the ambient field at transition was similar for fast and slow cool-downs confirming no impact of thermal currents. In all cases it was possible to observe a ÔjumpÕ at $T_\mathrm{c}=9.25$~K, which signals the transition into Meissner state and represents the flux expulsion from the cavity walls [see Fig.~\ref{fig:Meissner}]. The primary difference found was that fast cooling lead to a significant increase in the magnetic field right outside of the cavity walls while slow cooling lead to a much smaller change. This indicates that slow cool-down procedure seems to prevent flux from being expelled, leading to almost complete flux trapping, while a fast cool-down through $T_\mathrm{c}$ helps pushing efficiently the flux out of the superconductor. Higher amount of trapped flux correlated strongly with the observed higher residual resistance values for slow cooling. 

Exactly similar RF behavior was observed for the fine grain cavity with $R_\mathrm{res}(E_\mathrm{acc})$ curves shown in Fig.~\ref{fig:FGEP}. The magnetic field probe readings were also fully similar to those shown in Fig.~\ref{fig:SGEP}(b).

\subsection{Nitrogen doping}
We used 9-cell and 1-cell cavities prepared by the nitrogen doping procedure~\cite{Grassellino_SUST_2013} with the previously measured `anti Q slope' performance after regular fast cooling. 

\begin{figure*}[htb]
\includegraphics[width=\linewidth]{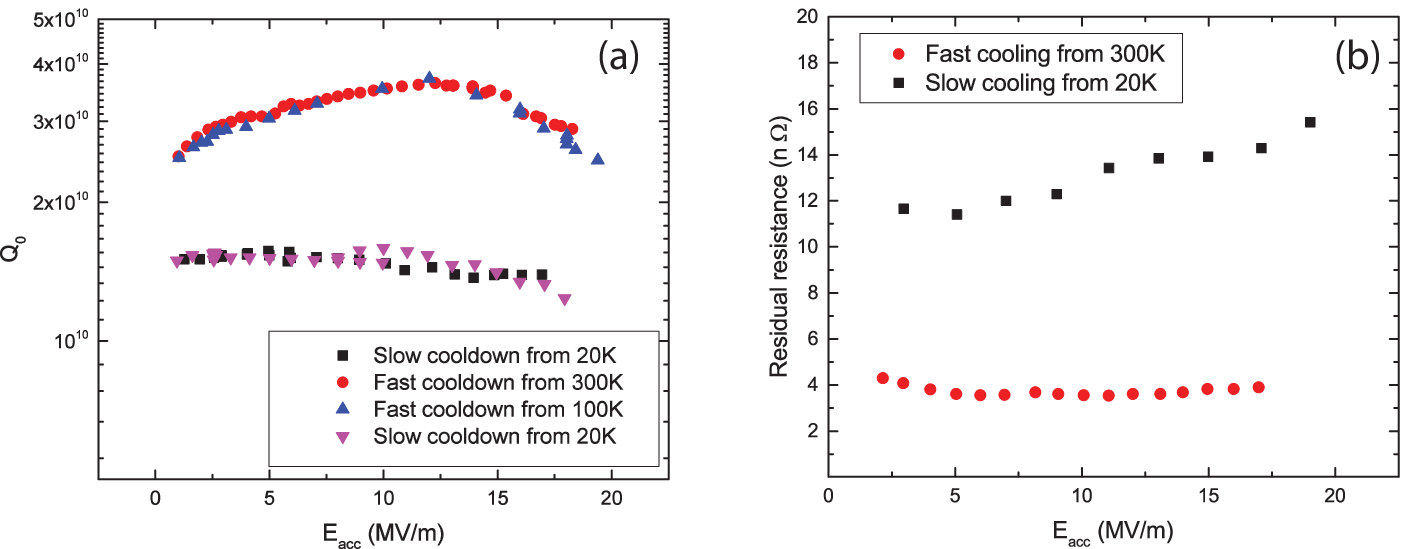}
\caption{\label{fig:TB9AES011}(a) $Q_0(E_\mathrm{acc})$ curves at 2~K for different cooling speeds measured on the 9-cell nitrogen doped cavity; (b) residual resistance for fast and slow cooling rates.}
\end{figure*}

No magnetic sensors were mounted in the 9-cell cavity experiment and only RF measurements were performed. We sequentially investigated fast and slow cooling procedures with the drastically different cooling rates and thermal gradients across the cavity described above with the results of the $Q_0(E_\mathrm{acc})$ measurements at $T=2$~K shown in Fig.~\ref{fig:TB9AES011}(a). 

\begin{figure*}[htb]
\includegraphics[width=\linewidth]{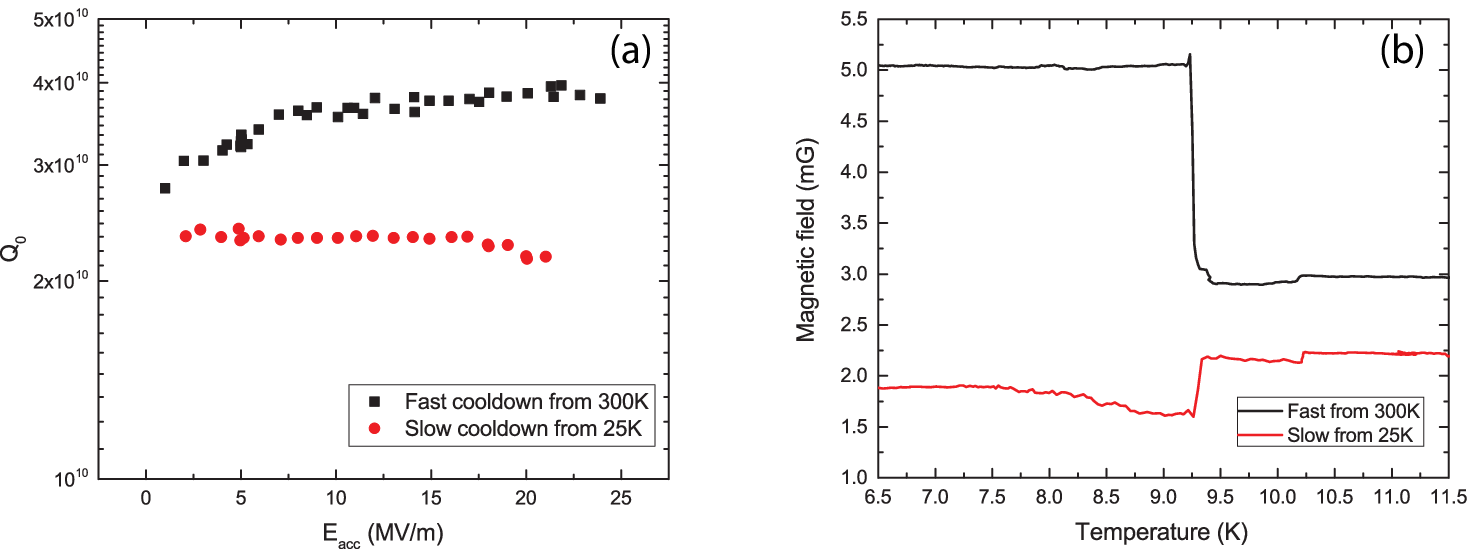}
\caption{\label{fig:TE1AES011}(a) $Q_0(E_\mathrm{acc})$ curves at 2~K for different cooling rates measured on the 1-cell nitrogen doped cavity; (b) corresponding magnetic field measurements around $T_\mathrm{c}$ (warm-up).}
\end{figure*}

A strong decrease in $Q_0$ was found in the case of the slow cooling procedure versus the fast. As in the case of EP, $Q_0(T)$ measurements at different $E_\mathrm{acc}$ revealed that the change in $Q_0$ arose from the increase in residual resistance as shown in Fig.~\ref{fig:TB9AES011}(b), while the BCS component remained unchanged. The low field residual resistance increased by about 8~n$\Omega$, and a stronger field dependence appeared, which lead to a total increase of $\gtrsim$10~n$\Omega$ in residual resistance at medium fields. 

In detail, the sequence of the tests was the following. The first RF test was performed after a typical fast cool-down from 300~K to 4.2~K. Second test followed a warm up to 20~K and slow cooldown through transition temperature similar to that in Fig.~\ref{fig:Cooling}(c). This resulted in a large residual resistance as shown in Fig.~\ref{fig:TB9AES011}(b) and a mediocre $Q_0$ vs. $E_\mathrm{acc}$ performance as illustrated in Fig.~\ref{fig:TB9AES011}(a). The cavity was then warmed up to 300 K and cooled down fast, similar to the cycle shown in Fig.~\ref{fig:Cooling}(b). As a result the performance recovered, reaching a residual resistance of $\sim$4~n$\Omega$ and a $Q_0\sim3\times10^{10}$ at medium fields at 2~K. The cavity was then warmed up to 100 K and held at 100 K for 8 hours to rule out the potential presence of hydrogen and Q-disease. From 100~K the cavity was then cooled down rapidly through $T_\mathrm{c}$. This resulted again in the good performance similar to the one after previous fast cool-down from 300~K. Cavity was then again warmed up to 20~K and the slow cool-down procedure was repeated, yielding again identical poor performance as in the previous slow cool-down. 

Next we studied the performance of a 1-cell nitrogen doped cavity with the 2~K results of $Q_0$($E_\mathrm{acc}$) shown in Fig.~\ref{fig:TE1AES011}(a). Similar to the 9-cell cavity, slow cooling lead to a significantly lower $Q_0$, again due to an increase in the residual resistance as was confirmed by lower temperature measurements. 

Magnetic field probes mounted on the cavity equator showed that ambient field at $T_\mathrm{c}$ was unaffected by the speed of the cool-down, but the drastic difference in the trapping efficiency between the slow and fast cooling was again observed - see Fig.~\ref{fig:TE1AES011}(b) where the warm-up data is shown for clarity. Also in this case the higher amount of trapped flux correlated well with the increase in $R_\mathrm{res}$ after slow cooling.

Notice that the increase in $R_\mathrm{res}$ for the 9-cell cavity is significantly higher than that for the 1-cell cavity. However, it may be simply a manifestation of the higher average magnetic fields sampled by the 9-cell in dewar with multiple potentially magnetic components. Indeed, later investigations showed that some of the holding fixtures possessed higher magnetic moments and could have lead to the observed difference. 

The magnitude of the $R_\mathrm{res}$ increase was nevertheless higher for nitrogen doped cavities as compared with the EP case. We will further discuss this point below.

\subsection{Electropolishing followed by 120$^\circ$C baking} 
\begin{figure*}[htb]
\includegraphics[width=\linewidth]{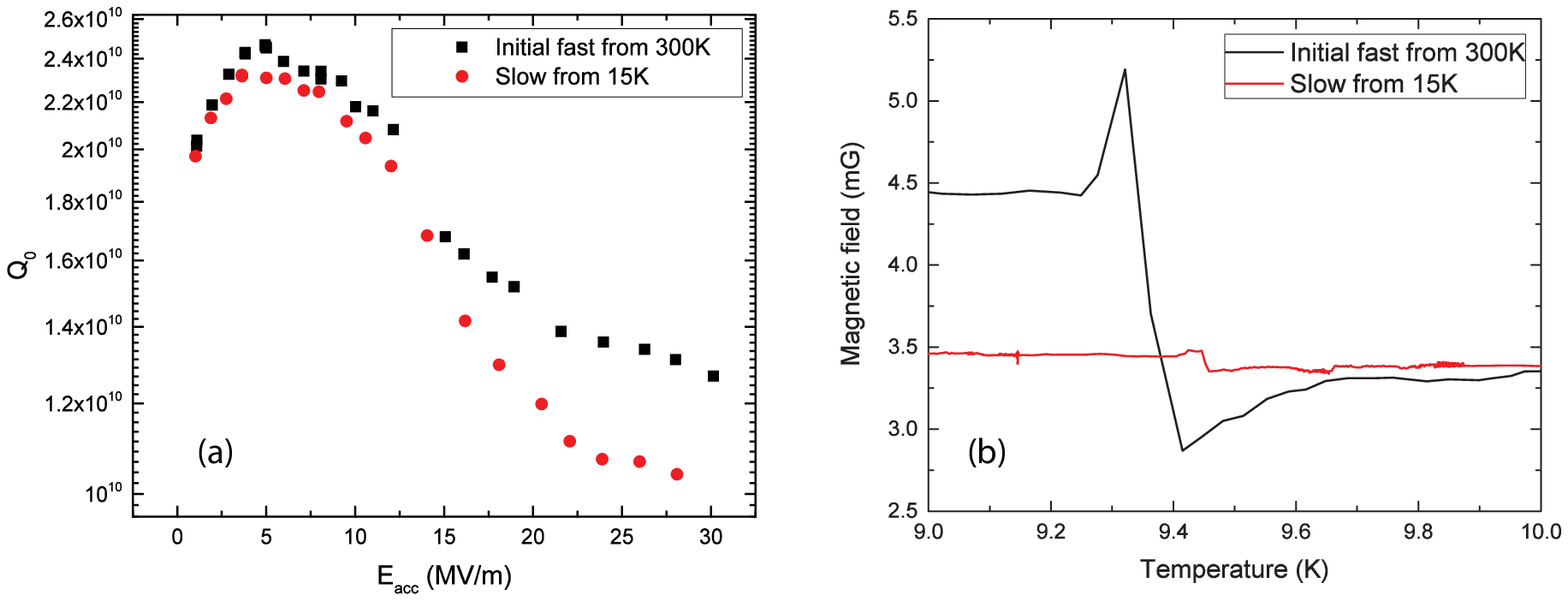}
\caption{\label{fig:TB9ACC012}(a) $Q_0(E_\mathrm{acc})$ curves at 2~K measured on the 9-cell cavity treated by the EP+120$^\circ$C baking process for fast and slow cooling procedures; (b) magnetic field at the equator recorded during cooling.}
\end{figure*}
\begin{figure*}[htb]
\includegraphics[width=\linewidth]{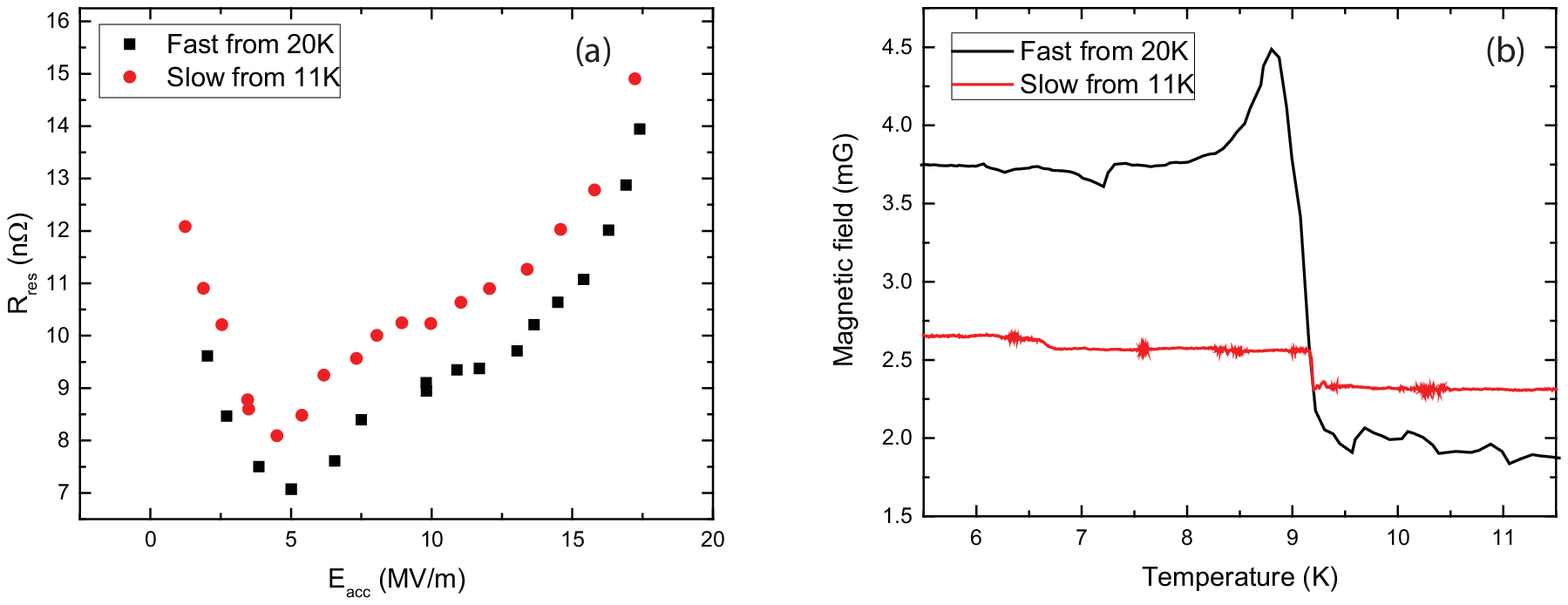}
\caption{\label{fig:FGEP120}(a) Residual resistance of the fine grain 1-cell EP+120$^\circ$C baked cavity after fast and slow cooling cycles; (b) corresponding magnetic field data at the equator (cool-down).}
\end{figure*}
In these experiments we investigated 9-cell and 1-cell cavities prepared by the standard ILC recipe, which consists of EP 120~$\mu$m + 800$^\circ$C baking for 3 hours + EP 20~$\mu$m + 120$^\circ$C baking for 48 hours.

The RF performance of the 9-cell was consistent with the previous findings, as shown in Fig.~\ref{fig:TB9ACC012}(a): fast cool-down lead at 2~K and medium fields to $Q_0 \sim 1.5\times10^{10}$, and slow to a lower $Q_0 \sim 1.2\times10^{10}$, again due to the residual resistance change. The same behavior was observed for the 1-cell cavity, for which the $R_\mathrm{res}(E_\mathrm{acc})$ is shown in Fig.~\ref{fig:FGEP120}(a) after fast cooling from 20~K, and after a slow cooling from 11~K. The additional residual resistance of about 1-2~n$\Omega$ emerged after a slow cooling.

Magnetic probes were mounted on the equator of the second cell from the top for the 9-cell test. The magnetic field ($\sim$3.5~mG) recorded by the probes [see Fig.~\ref{fig:TB9ACC012}(b)] right before transition was the same for both slow and fast cooling, while the suppression of the flux expulsion by slow cooling was again the only apparent difference correlating with the change in $Q_\mathrm{0}$ of the cavity. Similar behavior was registered also by the magnetic probe placed on the equator of the 1-cell during the corresponding test as shown in Fig.~\ref{fig:FGEP120}(b), which was again correlating with the observed change in $R_\mathrm{res}$ of the cavity. These data reconfirm that the mechanism behind the increase in residual resistance with slow cooling is lack of flux expulsion in contrast to strong and efficient flux expulsion obtained with fast cooling. 

Compared to the nitrogen doped and electropolished cavities the effect on the residual resistance is smallest in the 120$^\circ$C bake case, which will be further discussed below. 

\subsection{Effect of starting temperature on fast cooldown}
While cavity results above shown the same efficient flux expulsion upon fast cooling from 300, 100 and 20~K, it is important to understand if there exists a minimal starting temperature $T>T_\mathrm{c}$, which is required to maintain this efficiency. 

\begin{figure}[htb]
\includegraphics[width=\linewidth]{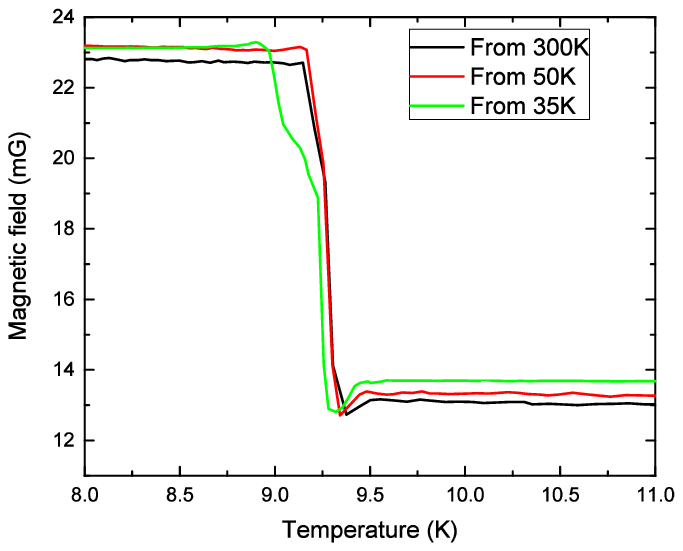}
\caption{\label{fig:Warmups}Trapped flux in the 9-cell nitrogen doped cavity after fast cooling from different starting temperatures.}
\end{figure}

To address this, we used another 9-cell nitrogen doped cavity and varied starting temperature of the cavity (and hence maximum temperature gradients) followed by multiple fast coolings through $T_\mathrm{c}$ recording the magnetic fields only (no RF test). In this experiment the ambient magnetic field at the probe location was also slightly higher ($\sim$12~mG). We found that the efficiency of flux expulsion was similar for starting temperatures of 300, 50, 35, 15, and 11~K - see Fig.~\ref{fig:Warmups} for examples of the magnetic field data taken during the warm-up. Temperature recording was interrupted during warm-ups after 15~K and 11~K cooling cycles but magnetic field recordings still showed the same jump at transition. This experiment suggests that cooling rate through $T_\mathrm{c}$ is what matters the most for an efficient expulsion rather than a starting temperature.

\section{Discussion}
\begin{figure*}[htb]
\includegraphics[width=\linewidth]{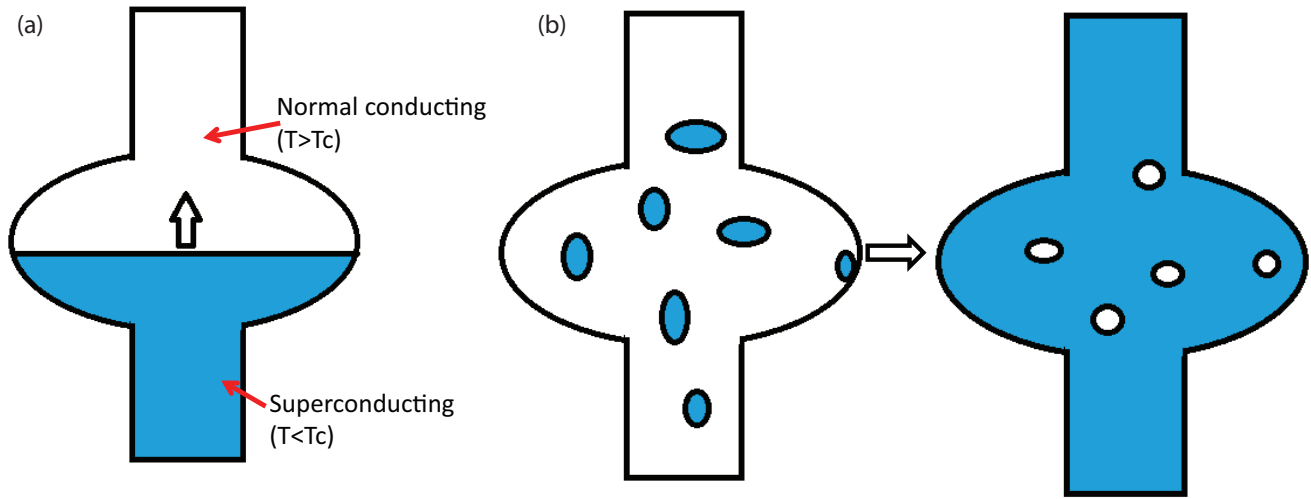}
\caption{\label{fig:Model}Schematic of the difference between the superconducting phase nucleation dynamics between the: (a) fast cool-down; (b) slow cool-downs}
\end{figure*}
\begin{figure}[htb]
\includegraphics[width=\linewidth]{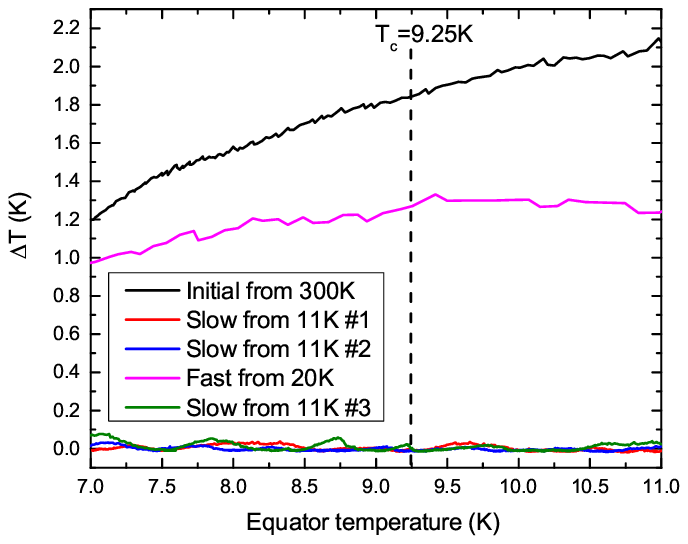}
\caption{\label{fig:Tgrad}Temperature difference between the equator and lower iris of the 1-cell cavity during different cooling procedures.}
\end{figure}

In our experiments flux magnetometers placed directly on cavity walls clearly showed that while the ambient field on the cavity surface was unaffected, the resulting residual resistance was strongly dependent upon the speed of the cool-down. Change in the field outside of the cavity allowed us to conclude that slow ($\leq$2-5~mK/sec) cooling through $T_\mathrm{c}$ leads to a much stronger external flux trapping in SRF niobium cavities than fast ($\gtrsim$30~mK/sec) cooling, which allows for much more efficient flux expulsion. This effect appears to be universal and independent on the surface treatment and grain size.

However, the same amount of trapped flux translated differently into RF losses for different surface treatments. This difference can be qualitatively understood based on the simplified model of the vortex dissipation $P_\mathrm{diss}$ under applied RF field, in which $P_\mathrm{diss}$ is proportional to the normal core surface area $A\propto\xi^2$ and the normal state resistivity $\rho_\mathrm{n}$. Since both the coherence length $\xi$ and $\rho_\mathrm{n}$ depend on the electron mean free path $\ell$ so does the $P_\mathrm{diss}$. It has been measured that $\ell \sim 2$~nm at the surface of a 120$^\circ$C baked cavity, $\ell \sim 40$~nm for nitrogen doped cavities~\cite{Romanenko_LEM_APL_2014, Grassellino_Talk_SRF_2013}, and $\ell \leq 100$~nm for unbaked EP cavities, hence the possible difference.

Cooling down a cavity is a complicated process and it is quite possible that the cooling rate is not an explicit parameter that determines amount of the trapped flux, but rather the parameter that determines a type of the cooling process which, in turn, changes the amount of flux trapped in the cavity.  Changing the cooling rate would also change few other parameters, most importantly evolution of the temperature distribution around the cavity.  

Nevertheless, based on our results the cooling through $T_\mathrm{c}$ has to be fast for in order to avoid the increase of the residual resistance due to trapped flux. Recent heat load measurements~\cite{Sekutowicz_private} in a cryomodule populated with 120$^\circ$C baked large grain 1.3~GHz dressed cavities seem to agree with these findings: the original fast cooldown lead to a high $Q_0$ (low heat load), then a slow cool-down through transition lead to significantly larger heat loads, and finally the low heat load (high $Q_0$) was recovered after warming up above and fast cooling back down through $T_\mathrm{c}$.

It should be emphasized that our findings do not contradict and may actually provide an alternative explanation/interpretation for the measurements on dressed cavities presented in Ref.~\onlinecite{Vogt_PRST_2013}. Indeed, in the HZB study the original cooling, which lead to the largest residual resistance, was by far the longest (18 hours) corresponding to the slowest rate of transition through $T_\mathrm{c}$. Thus, the magnetic flux expulsion should have been the least efficient, hence the increase in residual resistance. Corroborating this alternative interpretation of the experimental data is the fact that multiple vertical tests of dressed cavities performed at DESY~\cite{Info_DESY} showed no difference in the quality factors between bare and dressed cavities.

Other important experiments at Cornell University~\cite{Valles_IPAC_2013, Valles_PhD} showed that the residual resistance of the dressed cavity in the horizontal cryostat can be decreased by warming above $T_\mathrm{c}$ and slowly cooling back down, however the comparison is made with respect to the original cool-down, which is slow as well. Nevertheless, lack of increase in the surface resistance after fastest possible cooling from 100~K clearly suggests that the role of thermal gradients in those experiments is secondary (if any) and the same trapping efficiency dependence may be the main underlying mechanism of the observed changes. Measurements of the magnetic field at $T_\mathrm{c}$ and instantaneous (rather than average) cooling rate may provide further insight into the mechanisms at work.

To further clarify these physical mechanisms, we plan studies similar to ours but on dressed cavities. If thermocurrents will not be ruled out and actually do also play a role in dressed cavities, then the optimal cool-down procedure would be to cool down fast though transition but from a lower temperature, e.g. 20~K. 

If thermocurrents will be excluded, then attention should be paid exclusively to the rate of cooling through $T_\mathrm{c}$ from any starting temperature. 

At the moment, we offer two speculative interpretations for the observed trapping efficiency difference between fast and slow cool-downs.

The first one is the following. During a fast cool-down, liquid helium is being poured to the bottom of the warm dewar. Rising boiled-off helium gas sets a well defined temperature stratification around the cavity. In this situation the superconducting phase emerges at the bottom of the cavity and proceeds to sweep the cavity from the bottom to the top as shown in Fig.~\ref{fig:Model}(a). It looks plausible that the propagating phase boundary efficiently sweeps out the magnetic flux. During a slow cool-down the dewar and the cavity is cooled down by a cold gas produced by mixing in warm helium to the liquid helium inside the supply line. The temperature of this mixture is carefully controlled by varying the amount of the added warm gas. In a rough approximation the cavity can be considered to be isothermal and in thermal equilibrium with the cooling gas. In this scenario the superconducting phase would nucleate at multiple locations throughout the cavity. During the further cooling those interfaces can encircle some areas of normal phase as schematically shown in Fig.~\ref{fig:Model}(b). The magnetic flux contained in those 'islands', to get expelled, would need to pass through superconducting areas which is energetically unfavorable. This impediment would increase the amount of flux that gets trapped inside the superconductor.  The same qualitative behavior in single crystal and fine grain cavities despite the difference in bulk pinning seems to be in favor of this model as well. 

The second speculative possibility is that thermal gradients present during fast cooldown may exert a depinning force on the vortices, which counteracts trapping and helps pushing the flux out of the superconductor. In Fig.~\ref{fig:Tgrad} a temperature difference between the lower iris and equator of a fine grain 1-cell EP cavity is shown as a function of the equator temperature, which shows that temperature differences of the order of 1-2~K are present during the fast cooling. This translates into local temperature gradients of $\lesssim$0.4~K/cm, which may be high enough for fluxoid depinning around $T_\mathrm{c}$ leading to the more efficient flux expulsion in the case of fast cooling. During a very slow cool-down temperature gradients are unavoidably very small, no thermal depinning force is aiding the expulsion, and the efficiency of flux expulsion may decrease. 
 
We plan on addressing these models in further detail and explore other possibilities in future studies.

\section{Conclusion}
We have discovered a strong systematic effect of the cool-down rate through $T_\mathrm{c}$ on the ambient flux trapping efficiency of SRF niobium cavities. While the trapping effect itself is universal among surface treatments, the magnitude of the resulting changes in the residual resistance appears to be dependent on the surface treatment.

The reported findings are of primary importance for all the proposed high duty cycle accelerators based on SRF technology since preserving the low residual resistance allows to minimize their required operational power. The recommended operational cooling procedure in a cryomodule that currently emerges from these studies to minimize flux trapping is therefore to pass the 9.25~K transition temperature with a cooling rate $\gtrsim$30~mK/sec. Furthermore, since the effect is based on the trapping of the ambient magnetic field, increased magnetic shielding is an easy additional way to avoid the degradation in cases where only slow cooling can be performed.

\begin{acknowledgments}
We thank Peter Kneisel from JLab for providing the single crystal cavity for the studies. We acknowledge fruitful discussions with A. Crawford and help with cavity preparation and testing of A. Rowe, D. Bice, M. Wong, Y. Pischalnikov, B. Squires and all the FNAL cryogenic technical team. The work was partially supported by the DOE Office of Nuclear Physics. Fermilab is operated by Fermi Research Alliance, LLC under Contract No. DE-AC02-07CH11359 with the United States Department of Energy.
\end{acknowledgments}


\begin{thebibliography}{12}
\bibitem [{Inf()}]{Info_DESY}%
  \BibitemOpen
  \href@noop {} {}\bibinfo {note} {{Information can be found in DESY cavity
  database}}\BibitemShut {NoStop}%
\end{thebibliography}
%

\end{document}